\documentclass[twocolumn, tighten, times]{aastex631}
\usepackage{graphicx}
\usepackage{epstopdf}
\usepackage{multirow}
\usepackage{float} 
\usepackage{bm}

\usepackage{color}
\usepackage{soul}
\usepackage{enumitem}  
\usepackage{threeparttablex, tablefootnote}
\usepackage{booktabs}
\usepackage{ragged2e}
\usepackage{makecell}
\shorttitle{Classification of COSMOS}
\shortauthors{Fang. ET AL}
\graphicspath{{./}{figures/}}
\usepackage{multirow}
\usepackage{hhline}

\begin{document}

\title{Robustness Analysis of USmorph: II. Optimizing Feature Extraction, Dimensionality Reduction, and Clustering for Unsupervised Galaxy Morphology Classification}

\author[0000-0001-9694-2171]{Guanwen Fang}
\altaffiliation{Corresponding author: Guanwen Fang}
\affil{School of Physics and Astronomy, Anqing Normal University, Anqing 246011, China; 
\url{wen@mail.ustc.edu.cn}} 
\affil{Institute of Astronomy and Astrophysics, Anqing Normal University, Anqing 246133, China}

\author[0009-0004-5136-0951]{Xiaolei Yin}
\affil{School of Physics and Astronomy, Anqing Normal University, Anqing 246011, China; 
\url{wen@mail.ustc.edu.cn}} 
\affil{Institute of Astronomy and Astrophysics, Anqing Normal University, Anqing 246133, China}

\author[0000-0001-7707-5930]{Yirui Zheng}
\affil{School of Physics and Astronomy, Anqing Normal University, Anqing 246011, China;
\url{wen@mail.ustc.edu.cn}} 
\affil{Institute of Astronomy and Astrophysics, Anqing Normal University, Anqing 246133, China}

\author[0000-0001-8078-3428]{Zesen Lin}
\affil{Institute for Astrophysics, School of Physics, Zhengzhou University, Zhengzhou, 450001, China}

\author[0009-0004-0966-6439]{Shiwei Zhu}
\affil{School of Physics and Astronomy, Anqing Normal University, Anqing 246011, China; 
\url{wen@mail.ustc.edu.cn}} 
\affil{Institute of Astronomy and Astrophysics, Anqing Normal University, Anqing 246133, China}

\author[0000-0002-0846-7591]{Jie Song}
\affil{Department of Astronomy, University of Science and Technology of China, Hefei 230026, China; \url{xkong@ustc.edu.cn}} 
\affil{School of Astronomy and Space Science, University of Science and Technology of China, Hefei 230026, China}
\affil{Institute of Deep Space Sciences, Deep Space Exploration Laboratory, Hefei 230026, China}

\author[0000-0002-5133-2668]{Chichun Zhou}
\affil{School of Engineering, Dali University, Dali 671003, China}

\author[0000-0002-7660-2273]{Xu Kong}
\affil{Department of Astronomy, University of Science and Technology of China, Hefei 230026, China; \url{xkong@ustc.edu.cn}} 
\affil{School of Astronomy and Space Science, University of Science and Technology of China, Hefei 230026, China}
\affil{Institute of Deep Space Sciences, Deep Space Exploration Laboratory, Hefei 230026, China}

\begin{abstract}
We conduct a systematic robustness analysis of the unsupervised machine learning module within the hybrid framework \texttt{USmorph}. This module automatically discovers morphological structures from large-scale galaxy images, forming the foundation of the complete classification workflow. 
We evaluate five pre-trained models for feature extraction and identify an ImageNet-pretrained AlexNet as the most effective for capturing discriminative morphological features. UMAP is chosen for dimensionality reduction due to its optimal balance between preserving high-dimensional structure and computational efficiency. To enhance clustering stability, we propose a Bagging-based multi-cluster voting scheme, which significantly improves label consistency and cluster purity. We compare the convergence, scalability, and quality of five clustering algorithms, finding that the Bagging voting scheme has the best performance with the combination of K-means, Birch, and Agg. A bagging clustering number of $K=16$  is used to achieve the optimal balance between classification granularity and manual validation efficiency. Our tests show that: (1) the t-distributed stochastic neighbor embedding (t-SNE) reveals clear, compact cluster boundaries in low-dimensional space with strong feature separability; (2) the morphology classification results align with galaxy evolution theory, showing physically plausible distributions of different types in parameter space. 
These results demonstrate the technical robustness and scientific credibility of \texttt{USmorph}, establishing it as a reliable method for automated morphological classification in future large-scale surveys such as the China Space Station Telescope (CSST) mission.

\end{abstract}


\section{Introduction} \label{sec:1}

Galaxy morphology is a fundamental topic in astrophysics as it is tightly correlated with many physical properties of the galaxy (e.g., color, star formation rate, environment, mass, etc.)(e.g., \citealt{Kauffmann+2004,omand+2014, schawinski+2014, kawinwanichakij+2017,Gu+2018,su+2025}). Consequently, studying galaxy morphology helps us understand the formation and evolution of galaxies (see \citealt{Conselice+2014} for a review). 

Several methods have been used for describing galaxy morphology, with the most direct being visual inspection \citep{hubble+1926,Bergh+1976}. A notable example is the Galaxy Zoo project, which recruited a large number of volunteers to visually classify the morphology of nearly one million galaxies \citep{Lintott+2008, Lintott+2021}. 
Besides visual inspection, both parametric and non-parametric methods have been applied to galaxy morphology classification. 
Parametric measurements fit analytical functions to the luminosity profiles of galaxies and use the resulting parameters to describe galaxy morphology \citep{S+1963, Odewahn+2002, Balcells+2003}. 
Common parameters include the effective radius ($r_\mathrm{e}$), Sérsic index ($n$), axis ratio ($b/a$), and so on.  
In contrast, non-parametric methods compute statistical indicators directly from image pixels to characterize the morphological features of galaxies, such as the CAS system (Concentration $C$, Asymmetry $A$, and Smoothness $S$; \citealt{Conselice+2000, Conselice+2003}), the Gini-$M_{20}$ coefficient \citep{Lotz+2004,Lotz+2008}, and the MID system (Multimode, Intensity, Deviation; \citealt{Freeman+2013,Rodriguez_Gomez+2018}).

We are now in a new era of large-scale sky surveys, including the Sloan Digital Sky Survey (SDSS; \citealt{Stoughton+2002}), the Cosmic Evolution Survey (COSMOS; \citealt{Scoville+2007}), the Euclid space telescope (Euclid; \citealt{Euclid+2025}), and the upcoming China Space Station Telescope (CSST; \citealt{CSST+2025}). The next generation of sky surveys will produce an unprecedented volume of high-resolution galaxy images, a scale that renders traditional classification methods—including visual inspection, parametric, and non-parametric techniques—increasingly inadequate. Visual inspection is prohibitively time-consuming and difficult to scale, while traditional quantitative methods depend on predefined features that may not capture the full complexity of galaxy morphology. In contrast, machine learning methods can automatically learn discriminative features from images and identify complex non-linear patterns, offering a scalable, efficient, and consistent solution. They are thus particularly well-suited for automated galaxy morphology classification in the era of modern surveys.

Machine learning has been widely used in the study of galaxy morphology. In particular, convolutional neural networks (CNNs) have demonstrated considerable potential in galaxy morphology classification, as they possess the ability to automatically extract multi-level abstract features from image data and achieve precise modeling and classification of high-dimensional visual information through hierarchical feature learning. 
A series of studies have already applied CNNs to galaxy morphology classification (e.g., \citealt{Dieleman+2015, DominguezSanchez2018, Dickinson2018GalaxyZoo, walmsley2022galaxy}). 
Traditional supervised machine learning (SML) methods require substantial amounts of labeled data for training. Recent studies have shown that transfer learning and domain adaptation can partially alleviate this requirement by adapting pretrained models to new surveys with only a modest number of labeled targets (e.g., \citealt{Ciprijanovic+2023, Huertas-Company_Lanusse_2023}). Nevertheless, they still rely on some labeled data, and their performance remains sensitive to domain shifts—such as variations in point-spread function, imaging depth, and redshift—thereby limiting their general applicability across diverse surveys.
In contrast, unsupervised machine learning (UML) does not require pre-labeled samples and can directly perform clustering, dimensionality reduction, or feature learning on raw data, which helps reveal underlying patterns in the data.
UML is well-established as a powerful approach for efficient analysis of large-scale unlabeled galaxy datasets \citep[e.g.,][]{Hocking+2018,Martin+2020,cheng+2021,Tohill+2024}. 

However, UML methods still face challenges in practical applications. First, their performance is highly sensitive to the quality of the input data features. Raw high-dimensional data often contains significant noise, missing values, and irrelevant features, which can severely impair the effectiveness of the learned distance metrics. Furthermore, most UML methods employ a single clustering strategy, which often fails to capture true semantic relationships in data with complex manifold structures or imbalanced distributions. This limitation typically leads to unreasonable clustering and undermines the reliability of the clustering results.

In our previous research, \citet{Song+2024} proposed the \texttt{USmorph} framework for galaxy morphology classification, which includes feature extraction, unsupervised clustering, and supervised classification.
\texttt{USmorph} significantly reduces the reliance on labeled data when classifying new galaxy datasets. The framework includes the following processes: (1) Using a convolutional autoencoder (CAE; \citealt{Massey+2009}) to perform denoising and reconstruction of images; (2) Performing adaptive polar coordinate transformation (APCT; \citealt{Fang+2023}) on the denoised images to enhance the model's rotation invariance; (3) Using a Bagging-based multi-clustering voting method \citep{Zhou+2022} to cluster the extracted galaxy features; (4) Using labels obtained from clustering to train the CNN (GoogLeNet; \citealt{szegedy+2015}), and successfully classifying nearly 100,000 galaxies in the COSMOS field.

Although \texttt{USmorph} has demonstrated excellent performance in large-scale applications, its robustness under different configurations has not been systematically evaluated. 
Future sky surveys like those planned by the CSST will combine depth and breadth. It is therefore essential to ensure the \texttt{USmorph} framework can consistently produce stable and scientifically credible results. 
To this end, we conduct a systematic investigation with four key aspects.
First, while keeping the core components of the original \texttt{USmorph} framework unchanged (specifically, the CAE-based denoising stage and the APCT task design), we systematically compare the performance of five widely used pretrained encoders in the unsupervised morphology learning task, providing a quantitative assessment of encoder sensitivity. 
Second, based on AlexNet features, we conduct a comprehensive evaluation of four dimensionality-reduction methods and present quantitative comparisons in terms of separability and preservation of local/global structures. 
Third, we propose a Bagging-style multi-clustering voting strategy and demonstrate that this ensemble approach significantly improves both the purity and the stability of clustering assignments across multiple algorithms. 
Finally, we evaluate the impact of different choices of the number of clusters $K$ on clustering performance, intra-cluster consistency, and the efficiency of subsequent human labeling, thereby providing practical guidance for parameter settings in future large-scale surveys.

Throughout this paper, we use the AB magnitude system \citep{Oke+1983} and assume a \cite{Chabrier+2003} initial mass function and a standard flat $\Lambda$CDM cosmology with parameters $H_0 = 70$ km s$^{-1}$ Mpc$^{-1}$, $\Omega_m = 0.3$, and $\Omega_\Lambda = 0.7$.

\section{Data and Sample Selection } \label{sec:2}

\subsection{COSMOS}
The COSMOS survey covers an area of approximately 2 deg$^2$ and is specifically designed to investigate the interplay between galaxy evolution, star formation, active galactic nuclei (AGNs), dark matter, and large-scale structure across a redshift range of $0.5 < z < 6$ \citep{Scoville+2007,Weaver+2022}. 
In this study, we utilize high-resolution F814W-band images from the Hubble Space Telescope/Advanced Camera for Surveys (HST/ACS), covering an area of approximately 1.64 deg$^2$ within the COSMOS field. These image data were processed using the MultiDrizzle software package \citep{Koekemoer+2003}, resulting in images with a pixel scale of $0\farcs03$ and a 5$\sigma$ depth of 27.2 AB magnitude within a $0\farcs24$ aperture.

\subsection{COSMOS2020 Catalogue}

The COSMOS2020 ``Farmer'' galaxy catalog \citep {Weaver+2022} is one of the most comprehensive multi-wavelength photometric datasets currently available for extragalactic research, providing comprehensive photometric information across 35 bands from the ultraviolet to the near-infrared. 
Using this photometric dataset, \citet {Weaver+2022} fitted the spectral energy distributions (SEDs) of galaxies and estimated various physical properties of galaxies, including photometric redshifts, stellar masses, and star formation rates. 
To determine photometric redshifts, the catalog employed two different codes: {\tt EAZY} \citep {Bramme+2008} and {\tt LePhare} \citep {Ilbert+2006}. 
In this study, we adopt the redshifts obtained from {\tt LePhare} since the redshifts measured by {\tt LePhare} exhibit higher reliability within the relevant magnitude range \citep[see Figure 15 of][]{Weaver+2022}. 
Additionally, their analysis employed a series of dust extinction or attenuation curves, including the starburst attenuation curves proposed by \citet{Calzetti+2000}, the SMC dust extinction curves from \citet{Prevot+1984}, and two variants of the \citet{Calzetti+2000} attenuation law that include the 2175~\AA{} absorption feature. 
The final photometric redshift is defined as the median value obtained from the redshift likelihood distribution. Subsequently, the redshift is fixed at $z_{\text{LePh}}$, and the {\tt LePhare} fitting procedure is re-run to derive stellar mass and other related physical properties. Further details can be found in \citet {Laigle+2016} and \citet {Weaver+2022}.

\subsection{Sample Selection}

\begin{figure*}   
\includegraphics[width=1\textwidth]{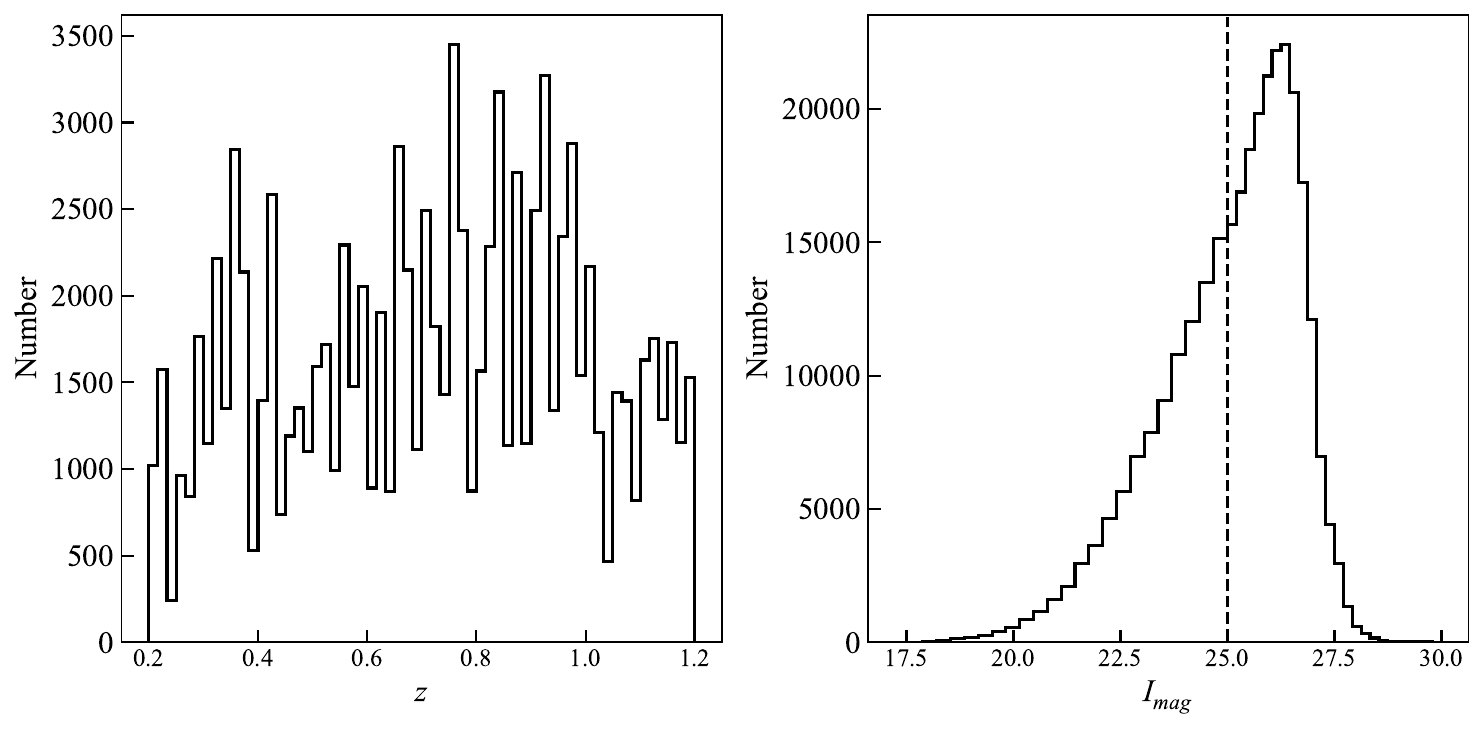}
\caption{
\textbf{Left:} the redshift distribution of the selected sample;
\textbf{Right:} the $I_{\rm{mag}}$ distribution of the COMOS2020 in the range of $0.2 < z < 1.2$ with the vertical dashed line indicating the brightness threshold $I_{\rm{mag}} < 25$ of the selected sample. }      
\label{fig:1}
\end{figure*}

In this work, we select galaxy samples from the COSMOS2020 catalog based on the following criteria:
\begin{enumerate}[label=(\arabic*)]

\item $\mathrm lp_{type} = 0$, to ensure that the samples we select are galaxies instead of stars; here, $\mathrm lp_{type}$ is the source-type flag in the catalog;
\item $I_{\rm{mag}} < 25$ mag, to exclude galaxies that are too faint to obtain reliable morphological measurements;
\item $\rm FLAG_{ COMBINE}=0$, to ensure that flux measurements are not affected by bright stars and the object is located at the center of the image, thereby guaranteeing the reliability of photometric redshift and mass estimates;
\item $0.2 < z < 1.2$, to ensure the measurements of galaxy morphology in the rest-frame optical band;
\item High-quality source images with a signal-to-noise ratio (S/N) greater than 5 and no abnormal pixels, to ensure the high quality of source images.
\end{enumerate}
The selected sample contains 99,806 galaxies, and their distribution in I-band magnitude and redshift is shown in Figure~\ref {fig:1}.

\section{The Unsupervised Method for Morphological Classification} \label{sec:3}

\begin{figure*}    
\includegraphics[width=1\textwidth]{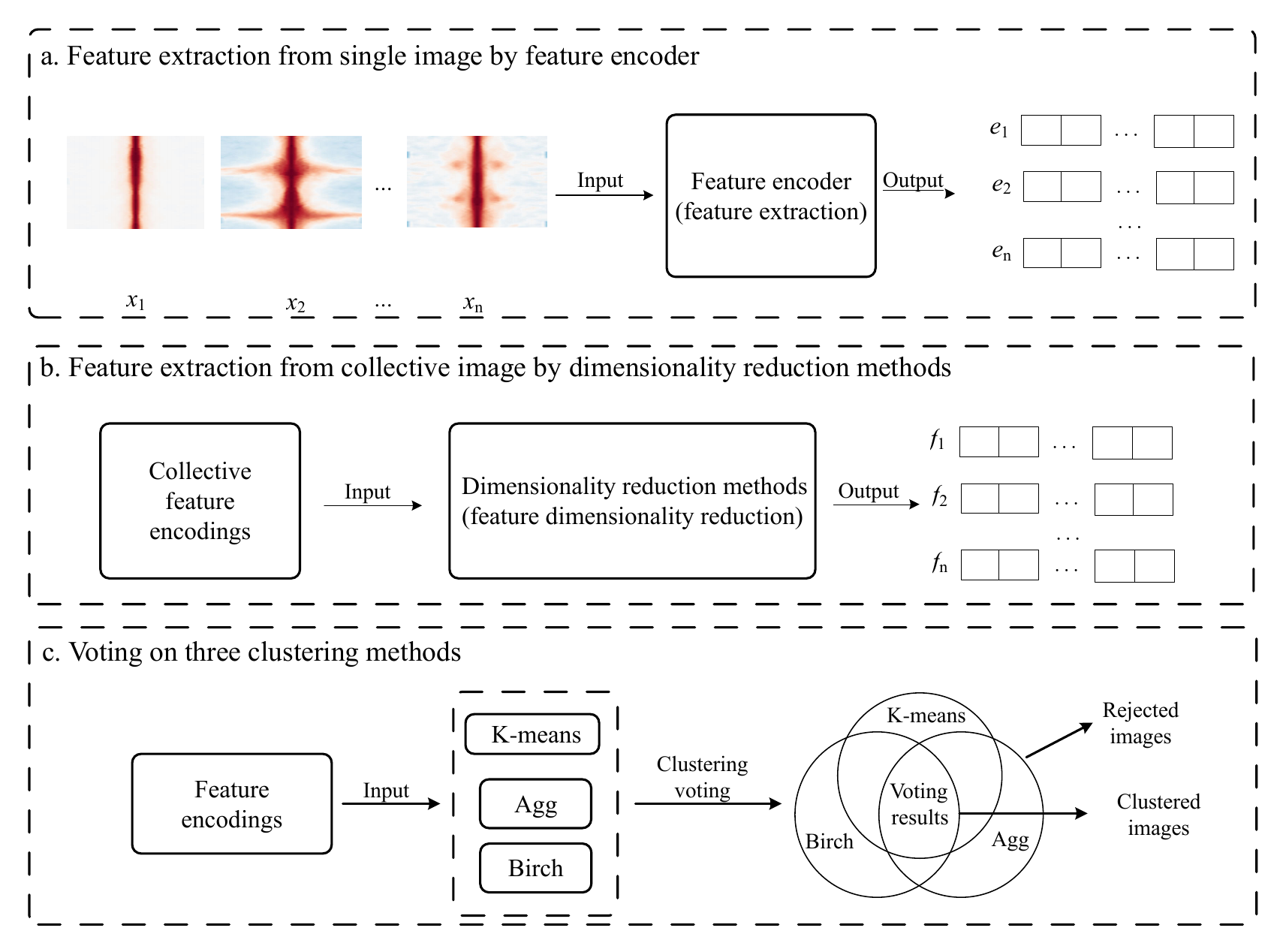}
\caption{Flowchart of the unsupervised galaxy morphology classification pipeline of the \texttt{USmorph} framework, comprising three main stages: feature extraction (Panel \textbf{(a)}), dimensionality reduction (Panel \textbf{(b)}), and unsupervised clustering (Panel \textbf{(c)}).}      
\label{fig:2}
\end{figure*}

This section presents extensive experiments conducted to determine the optimal configuration specifically for the UML component within the \texttt{USmorph} framework under current conditions. As illustrated in Figure \ref{fig:2}, the UML clustering process primarily involves the following steps: First, preprocessed galaxy images are input into a pre-trained model for feature encoding to obtain high-dimensional feature representations. Second, dimensionality reduction techniques (e.g., Principal Component Analysis, Uniform Manifold Approximation, Projection, Random Projection and Mean Pooling over Blocks) are applied to compress these features into a low-dimensional space, reducing redundancy and improving computational efficiency for subsequent steps(see Section~\ref{sec:3.2.2} for details). Next, multiple clustering algorithms are executed on the reduced-dimensional feature space to group samples. Their clustering results are then fused through a voting mechanism to achieve a more stable and reliable partition. Finally, the resulting clusters undergo visual inspection and summarization, and are mapped to common galaxy morphological categories to generate the final morphological classification results.  Based on this workflow, we evaluate different feature encoders and dimensionality reduction techniques. 
We further test combinations of three clustering algorithms for the Bagging-based voting model and determine the optimal value for the number of clusters ($K$).

\subsection{Data Preprocessing}

\begin{figure*}    
\includegraphics[width=1\textwidth]{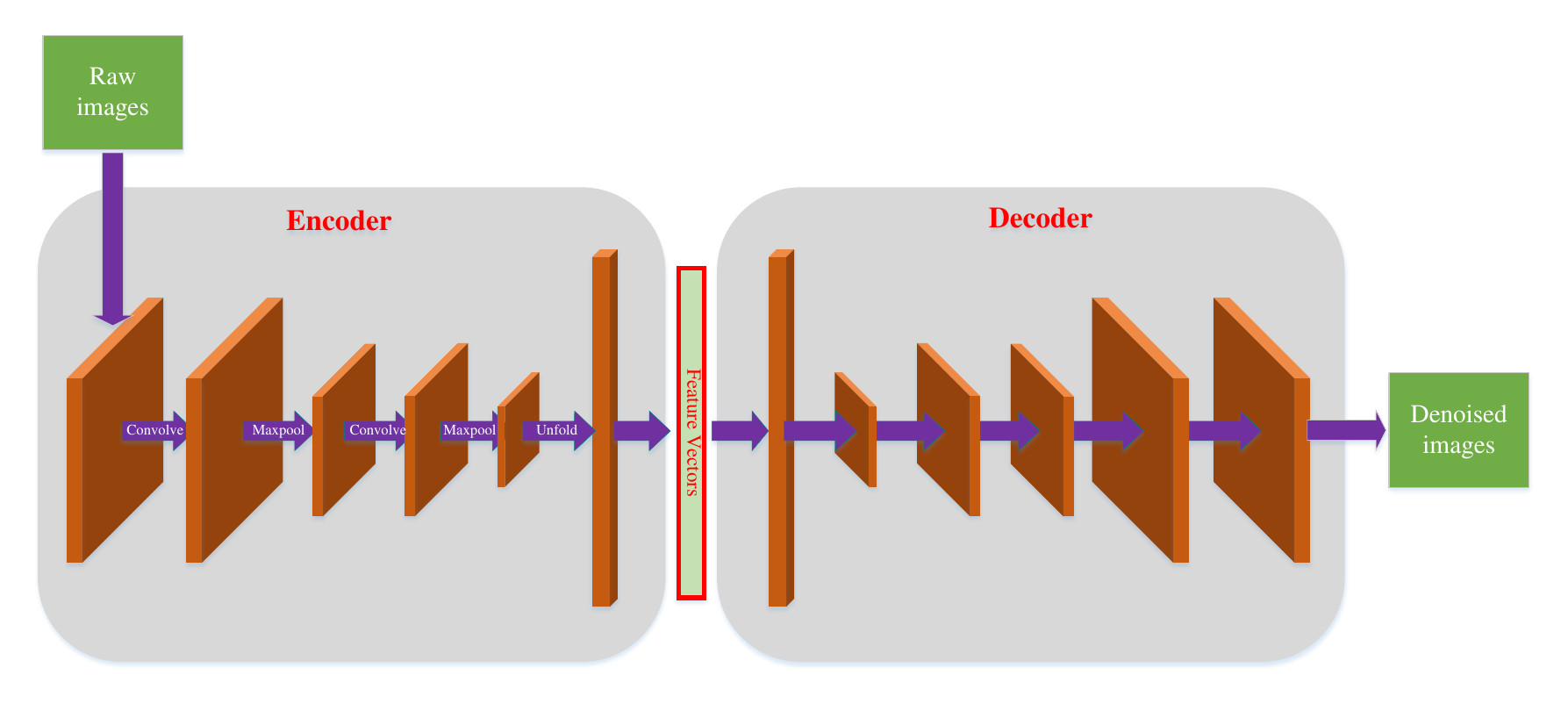}
\caption{Schematic illustration of the CAE architecture. The left part is the encoder, and the right part is the decoder. The original image is fed as input, and a denoised image is reconstructed at the output after processing through the encoder and decoder.}      
\label{fig:3}
\end{figure*}

To enhance the stability and accuracy of the clustering model, we perform data preprocessing on galaxy images to reduce image noise and improve rotational invariance.
Figure~\ref{fig:3} shows the employed denoising framework that is based on a CAE \citep{Massey+2010}. The framework first extracts latent features from raw galaxy images through a series of convolutional and pooling operations and then reconstructs these features with deconvolution and upsampling operations to generate the final denoised images. The CAE configuration used here was carefully validated through comparative experiments to ensure that the denoising process preserves galaxy structural features for the downstream analysis.
The architectural parameters and a comprehensive description of the CAE-based denoising framework are provided in \citet{Zhu+2025}.

After denoising the raw images, we need to further improve rotational invariance.
Previous studies have shown that spatial rotation transformations can lead to incorrect classification of galaxy morphological types by the model, significantly reducing its performance \citep{Dieleman+2015, Yao+2019}. 
To overcome this limitation, we adopt the APCT technique proposed by \cite{Fang+2023}, whose study demonstrated that APCT significantly enhances the robustness of convolutional neural networks to orientation variations, thereby improving classification accuracy in scenarios requiring rotational invariance.
The APCT technique first defines the initial polar axis based on the pixels with extreme values (i.e., maximum and minimum brightness) within the image. This polar axis is then systematically rotated counterclockwise in increments of 0.05 radians. At each rotation angle, pixel intensities along the current polar axis direction are integrated via polar coordinate projection. A mirroring operation is then applied to enhance centrally symmetric features. 
This coordinate system transformation enhances the representation of galaxy morphological structures while improving the model's rotational invariance.

\begin{figure*}    
\includegraphics[width=1\textwidth]{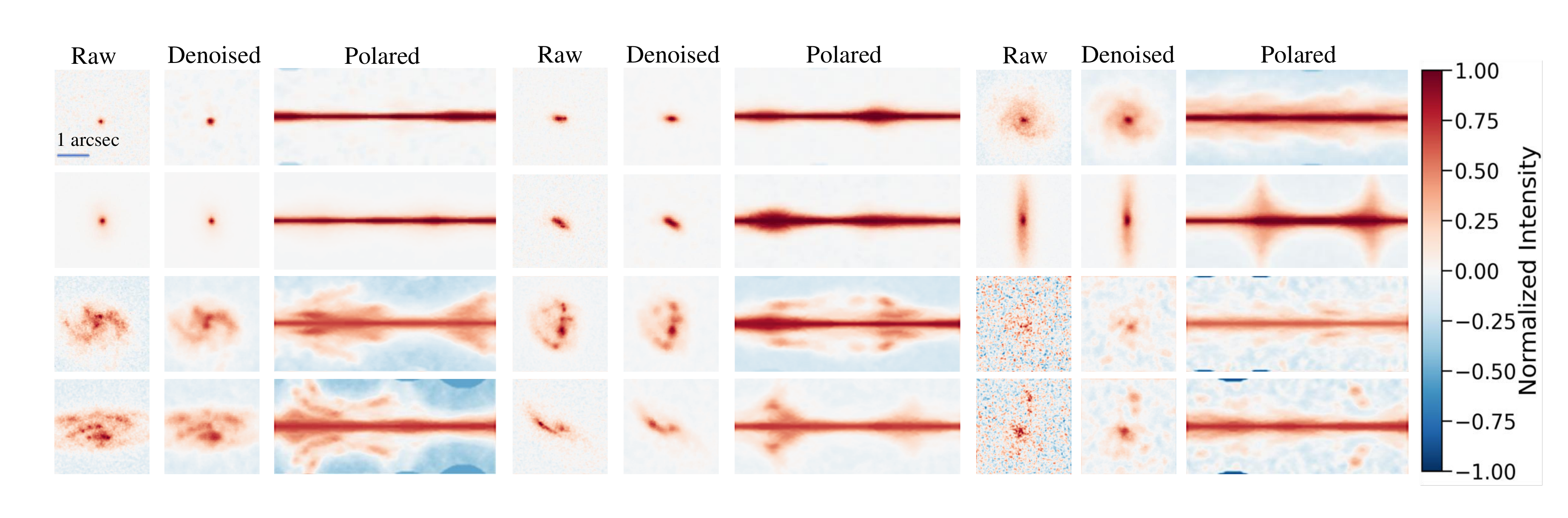}
\caption{Six image sets that demonstrate image preprocessing steps. Each set contains two galaxies of the same classification category. In each set, the left, center, and right panels show the original images in the rest-frame optical band, post-CAE-based denoised images, and the images after polar coordinate expansion, respectively. The blue bar in the first panel indicates an angular scale of $1^{\prime\prime}$ ($\approx 33$ pixels).}      
\label{fig:4}
\end{figure*}

Figure~\ref{fig:4} shows the effects of the CAE denoising framework and the APCT technique with six image sets.
Each set consists of three components: the original image (left panels), the denoised reconstructed image generated by CAE (central panels), and the polar coordinate transformation output by ACPT (right panels). The significant differences between the original image and the denoised image demonstrate that the CAE architecture effectively removes noise from the image while preserving its key morphological features. 
Similarly, the right panels show that APCT processing effectively highlights structures like spiral arms.
The significant differences between the original images and the preprocessed ones demonstrate the effectiveness and necessity of noise reduction and rotational invariance improvement.

\subsection{UML Clustering Process}

\subsubsection{Feature Encoder}

Large-scale pre-trained models have achieved remarkable advances in computer vision, establishing themselves as a technical cornerstone for image understanding and feature learning \citep{Krizhevsky+2012, he+2016deep,Dosovitskiy+2020}. 
These models are typically pre-trained on large-scale image datasets (such as ImageNet and JFT-300M) \citep{Deng+2009, sun+2017} using either supervised or self-supervised learning strategies \citep{He+2020, Cheng+2020}, leveraging deep neural network architectures (e.g., ResNet and Vision Transformer; \citealt{he+2016deep,Dosovitskiy+2020}) to learn general-purpose visual representations. 
After sufficient training, intermediate layers of the model can efficiently extract features, transforming raw images into high-dimensional feature vectors rich in semantic information. These features are widely used in various downstream tasks, including image classification, object detection, image retrieval, and medical image analysis \citep{girshick+2014rich, Russakovsky+2015, litjens+2017}.
It offers several advantages to extract image features with pre-trained models. During pre-training, the model captures hierarchical visual features ranging from low-level edges and textures to high-level semantic concepts \citep{Zeiler+2014}, effectively reflecting the essential structure and content of images. Additionally, using pre-trained models reduces the reliance on extensive annotated data. Their capability of transfer learning enables reliable feature extraction, which allows downstream tasks to achieve good performance even when labeled samples in the target domain are limited \citep{yosinski+2014}. With growing model capacity and scale of training data, the generalization ability and cross-domain adaptability of the learned features are further enhanced.

\begin{figure*}    
\includegraphics[width=0.9\textwidth]{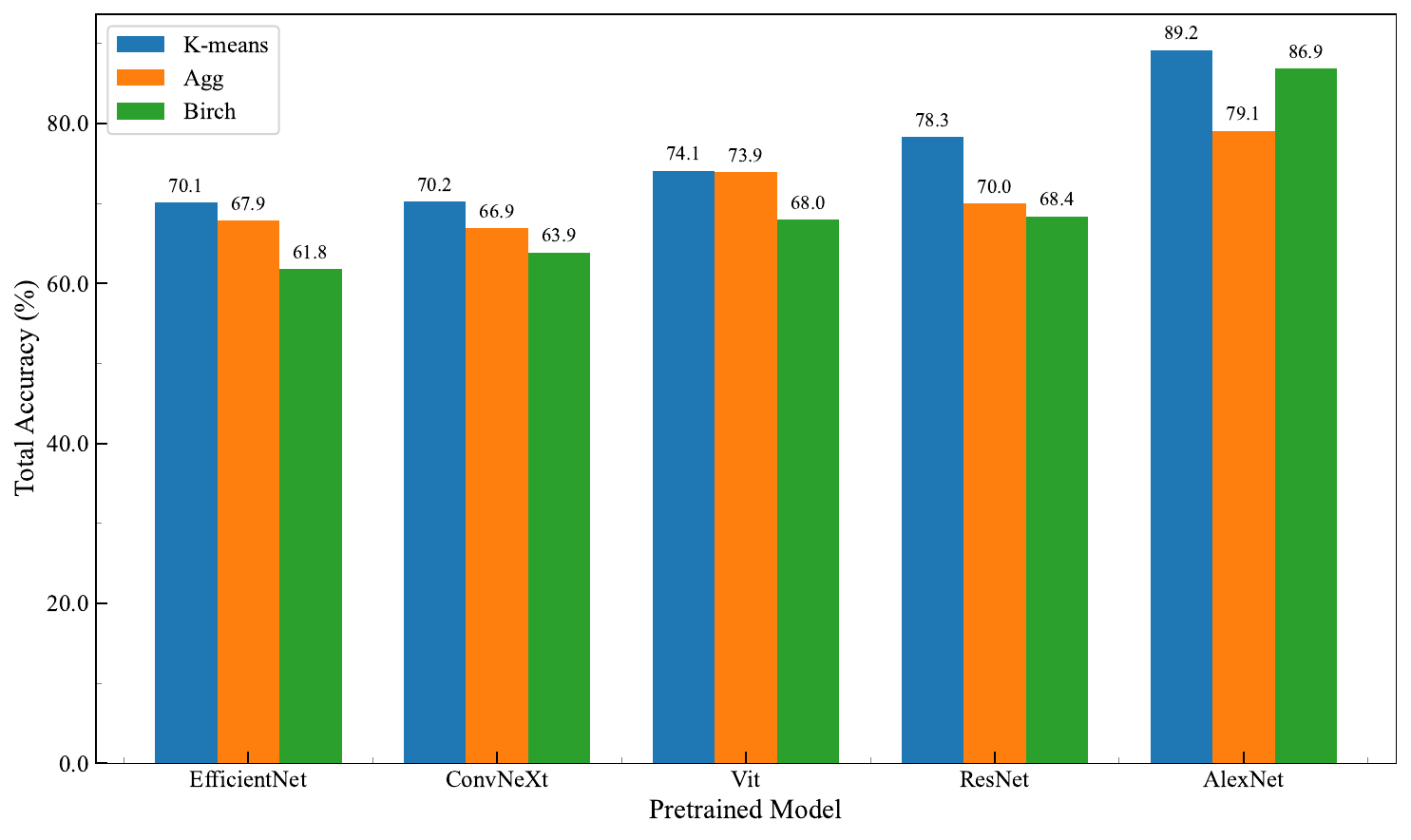}
\caption{Comparison of total accuracy for different pretrained models and clustering methods. }
\label{fig:5}
\end{figure*}

In this work, we evaluate five pre-trained models for their performance in feature extraction, including AlexNet \citep{Krizhevsky+2012}, ResNet \citep{he+2016deep}, EfficientNet \citep{Tan+2019}, ViT \citep{Dosovitskiy+2020}, and ConvNext \citep{liu+2022}. Our experiments confirm that AlexNet achieves the best performance on the galaxy morphology classification task, as shown in Figure~\ref{fig:5}. 
In this experiment, we use different pre-trained models as feature extractors to encode the preprocessed galaxy images. With these extracted features, we apply three clustering algorithms to generate cluster assignments, which are treated as pseudo-labels. We then train a supervised classifier (GoogLeNet; \citealt{szegedy+2015}) on the preprocessed galaxy images using these pseudo-labels, with a train/test split of 9:1 \citep{Fang+2023}, and report the classification accuracy on the held-out test set. The accuracy is defined as
\begin{equation}
\mathrm{Acc}
=\frac{1}{N}\sum_{i=1}^{N}\mathbb{I}(\hat{y}_i = y_i),
\label{eq:acc}
\end{equation}
where $N$ is the number of evaluated samples, $y_i \in \{1,\dots,K\}$ is the ground-truth class label of the $i$-th sample, $\hat{y}_i$ is the predicted class label, $K$ is the number of classes, and $\mathbb{I}(\cdot)$ is the indicator function, which equals $1$ if the condition holds and $0$ otherwise.
This accuracy is used as a proxy metric to evaluate the effectiveness of feature representations produced by different pre-trained models.

All five pretrained models considered in this work were initialized with weights learned from the same large-scale natural image dataset (ImageNet). Therefore, the performance differences among them should be attributed to the architectural characteristics rather than differences in pretraining data. AlexNet benefits from its shallow convolutional layers that effectively capture low-level patterns. This aligns well with the fact that galaxy morphology is defined by such patterns, including global structure, symmetry, and local geometric features like spiral arms and core distribution. Meanwhile, AlexNet's lower model complexity enables stronger generalization on datasets of the ten-thousand-scale.
Additionally, the standard convolutional operations of AlexNet are well-suited for smooth, symmetric astronomical images, whereas the patch-based attention mechanism of ViT may introduce spatial fragmentation when processing continuous galaxy structures.
The better alignment with the target data distribution and task essence supports AlexNet's best performance among the five models, despite its lack of state-of-the-art architectural features.

\subsubsection{Feature Dimensionality Reduction}\label{sec:3.2.2}

The feature representations extracted by deep pre-trained models from preprocessed images are often high-dimensional.
While these high-dimensional features contain rich semantic information, their dimensionality itself poses significant challenges for downstream clustering tasks.
With higher dimensionality, data becomes more sparse in high-dimensional space. It undermines the discriminative ability of traditional distance metrics (e.g., Euclidean distance), making it more difficult for clustering algorithms to accurately capture similarity relationships among samples. 
This ``curse of dimensionality'' significantly increases computational and storage costs, thereby reducing clustering efficiency \citep{Aggarwal+2001surprising,Kriegel+2009clustering}. This issue is particularly pronounced when handling large-scale datasets. Additionally, high-dimensional features often contain redundant information or noise, which may interfere with the clustering process and lead to unstable results or degraded performance \citep{qu+2023}.
It is essential to reduce dimensionality before clustering.

Proper dimensionality reduction techniques can project data into a lower-dimensional space while retaining its intrinsic structure and critical information.
They effectively mitigate the curse of dimensionality, improve computational efficiency, and enhance the robustness of clustering algorithms. 
By compressing dimensions and removing redundancy, these methods reveal the intrinsic clustering structure in a lower-dimensional space, thereby providing more compact and discriminative representations for subsequent clustering analysis.
Thus, appropriate dimensionality reduction serves as a crucial bridge connecting deep feature extraction with efficient clustering.

Several dimensionality-reduction techniques are commonly used. In particular, Principal Component Analysis (PCA; \citealt{MACKIEWICZ+1993}) and Uniform Manifold Approximation and Projection (UMAP; \citealt{McInnes+2018}) are widely adopted.
PCA is a linear dimensionality reduction technique.
It identifies orthogonal directions of maximum variance (principal components) by computing eigenvalues and eigenvectors. PCA then projects the data onto a lower-dimensional subspace defined by these principal components. Such a process preserves the global structure of the data while enabling effective data compression and noise reduction. 
In galactic morphology classification, PCA compresses high-dimensional galaxy image features into a lower-dimensional space. It removes redundancy while preserving key morphological information, thereby improving the accuracy of classification and clustering. 
In contrast, UMAP is a nonlinear technique that effectively preserves both local and global data structures. By constructing a topological representation of the high-dimensional data, UMAP optimizes the preservation of local neighborhoods while maintaining broader global relationships. This makes it especially effective for capturing complex, nonlinear patterns in high-dimensional data such as galaxy images. In galaxy morphology classification, UMAP effectively groups similar galaxy samples by capturing subtle local similarities between different galaxy types. Therefore, UMAP enhances cluster separation and improves downstream clustering or classification performance \citep{Fang+2026updated}. 
To evaluate the effectiveness of UMAP and PCA in our \texttt{USmorph} framework, we further test two additional classical dimensionality-reduction methods: Random Projection (RA; \citealt{Bingham+2001random}) and Mean Pooling over Blocks (BM; \citealt{Boureau+2010theoretical}). RA multiplies the feature vectors with a sparse random projection matrix whose shape is set to match the desired output dimension.
It reduces the data dimensions while approximately preserving pairwise distances at minimal computational cost. 
In contrast, BM is a simple and fast method that partitions the vector into contiguous segments (``blocks'') corresponding to the target dimensionality and takes the mean within each block.

Our experiments show that UMAP yields the highest accuracy for galaxy morphology classification among the four methods. To ensure a fair comparison, we extract features using the same pre-trained AlexNet encoder and evaluate different dimensionality reduction algorithms under identical clustering settings. The results indicate that, compared with the other dimensionality reduction methods, UMAP provides a more effective low-dimensional embedding, leading to higher classification accuracy and improved computational efficiency. Detailed results are summarized in Table~\ref{dimensionality reduction method}, where the reported accuracy is computed with Equation~(\ref{eq:acc}).

\begin{table}[htbp]
\centering
\caption{Clustering accuracy of AlexNet features under different dimensionality reduction and clustering methods.}\label{dimensionality reduction method}
\small
\renewcommand{\arraystretch}{1}
\setlength{\tabcolsep}{10pt}
\begin{tabular}{|c|c|c|c|}
\hline
\textbf{Model} & \textbf{DR Method} & \textbf{Clust.} & \textbf{Acc (\%)} \\
\hline
\multirow{12}{*}{AlexNet}
& \multirow{3}{*}{RA}    & K-means & $84.7 $ \\ \hhline{|~|~|-|-|}
&                         & Agg    & $78.1 $  \\ \hhline{|~|~|-|-|}
&                         & Birch  & $79.5 $  \\ \hhline{|~|~|~|-|}
\cline{2-4}
& \multirow{3}{*}{BM}     & K-means & $86.3 $  \\ \hhline{|~|~|-|-|}
&                         & Agg    & $77.9 $ \\ \hhline{|~|~|-|-|}
&                         & Birch  & $79.6 $ \\ \hhline{|~|~|~|-|}
\cline{2-4}
& \multirow{3}{*}{PCA}    & K-means & $87.7 $  \\ \hhline{|~|~|-|-|}
&                         & Agg    & $85.8 $  \\ \hhline{|~|~|-|-|}
&                         & Birch  & $80.3 $  \\ \hhline{|~|~|~|-|}
\cline{2-4}
& \multirow{3}{*}{UMAP} & K-means & $88.6 $  \\ \hhline{|~|~|-|-|}
&                         & Agg    & $86.5 $  \\ \hhline{|~|~|-|-|}
&                         & Birch  & $87.2 $  \\ \hline
\end{tabular}
\par\vspace{2mm}
\noindent \footnotesize\justifying{\textit{Note:} The second column lists different dimensionality reduction methods, and the third column lists different clustering algorithms.}
\label{tab:alexnet_clustering}
\end{table}

An appropriate choice of dimensionality is crucial when applying linear dimensionality reduction methods. An optimal dimension preserves the majority of data variance, eliminates redundancy, and captures essential features effectively. Common approaches for determining the optimal dimension include analyzing the cumulative explained variance or the information ratio. In this work, we employ the \textit{elbow method} to identify the optimal number of dimensions. Specifically, we first compute the covariance matrix of the data and perform eigenvalue decomposition to project the data onto a new feature space spanned by the principal components. We then extract the eigenvalues corresponding to each principal component and compute the \textit{information ratio}, defined as the ratio between consecutive eigenvalues. 
The information ratio at component $k$ is computed as:
\begin{equation}
    \text{Information Ratio}_k = \frac{\lambda_{k+1}}{\lambda_{k}},
\end{equation}
where $\lambda_k$ denotes the eigenvalue of the $k$-th principal component.
Information ratio serves as an indicator of the relative importance between adjacent components and helps identify the ``elbow'' point, the dimension at which the marginal gain in information drops significantly. The dimension corresponding to this turning point is derived as the optimal reduced dimension. 

\begin{figure}    
\includegraphics[width=0.49\textwidth]{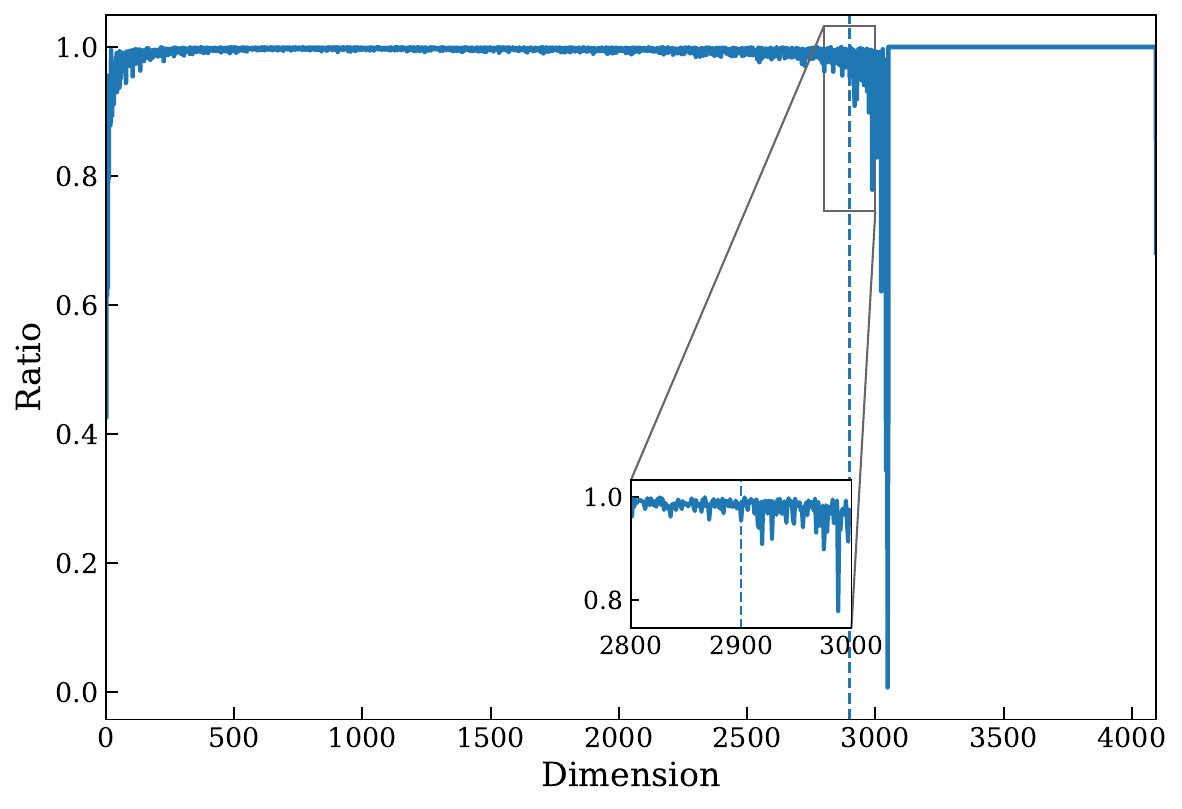}
\caption{The information ratio as a function of dimension with the PCA method. The ratio starts to drop significantly after 2900 dimensions, indicating that the main features are concentrated within the first 2900 dimensions. Therefore, we choose to reduce the dimensionality to 2900.}
\label{fig:6}
\end{figure}

We compute the information ratio as a function of dimension with the PCA method and plot the results in Figure~\ref{fig:6}.  
The result shows a distinct ``elbow'' point, after which the ratio starts to drop significantly.
This indicates that the principal features are concentrated within the first 2900 dimensions, which we therefore select as the optimal dimensionality for linear reduction.

The Davies-Bouldin Index (DB Index; \citealt{Davies+1979}) is a widely used internal metric for evaluating clustering quality. 
As an unsupervised measure, it operates independently of external class labels and assesses the validity of clustering results by quantifying the ratio between intra-cluster compactness and inter-cluster separation. 
The DB Index is defined as:
\begin{equation}
    \mathit{DB} = \frac{1}{N} \sum_{i=1}^{N} \max_{j \neq i} \left( \frac{S_i + S_j}{d(c_i, c_j)} \right),
\end{equation}
where $N$ is the total number of clusters;
$S_i$  ($S_j$) represents the cohesiveness of clusters $i$ ($j$), defined as the average distance between its members and its centroid $c_i$ ($c_j$); $d(c_i, c_j)$ denotes the distance between the centroids $c_i$  and $c_j$.
The DB Index is particularly effective for evaluating clusters with convex geometries \citep{Halkidi+2001}.
\textit{Lower} DB Index values indicate more compact and separated clusters, implying better clustering performance.
To balance dimensionality reduction effectiveness and computational efficiency in the UMAP dimensionality reduction process, we compute the DB Index across a range of dimensions from 50 to 500 in steps of 50.
The index is minimized at 300 dimensions; consequently, we use 300 dimensions for downstream analysis. This approach is simple, parameter-light, but provides a reliable, data-driven criterion for determining the optimal reduced dimensionality.

\subsubsection{Clustering Methods}

To improve the stability and robustness of the hybrid clustering framework for galaxy morphology classification, we systematically evaluate several clustering algorithms and analyze their influence on classification performance. 
As an unsupervised learning paradigm, clustering partitions data samples by their intrinsic similarities.
It aims to maximize cohesion within clusters while maintaining clear separation between different clusters, which is essential for effectively distinguishing different morphological types of galaxies, such as ellipticals, spirals, and irregulars, from their high-dimensional feature representations.
However, the performance of clustering algorithms is highly sensitive to the underlying data distribution, the structure of the feature space, and the intrinsic complexity of the morphological classes.

We compare five commonly used clustering approaches: K-means \citep{Hartigan+1979}, Agglomerative Clustering (Agg; \citealt{Murtagh+1983, Murtagh+2014}), Birch \citep{Zhang+1996}, DBSCAN \citep{Ester+1996, Campello+2013}, and Spectral Clustering \citep{Ng+2001spectral}. 
Three of them exhibit distinct strengths. K-means provides fast and efficient partitioning based on centroid distances, performing well on approximately spherical and well-separated clusters. Birch demonstrates strong scalability for large astronomical datasets by constructing a compact Clustering Feature (CF) tree with low memory overhead \citep{Zhang+1996,Yang2022data}. Agglomerative Clustering effectively captures hierarchical patterns within the data, which is particularly valuable for identifying nested or multi-scale galaxy morphologies \citep{cheng+2021,Yu+2022hierarchical}. 

To address the limitations of individual algorithms, particularly in robustly classifying ambiguous or transitional morphologies like merging systems or low-SNR objects, we employ a voting-based ensemble strategy that integrates the clustering results of K-means, Birch, and Agg. 
This consensus approach effectively reduces the bias and instability associated with any single method, leading to a significant improvement in clustering purity and morphological consistency. Although DBSCAN and Spectral Clustering can recover non-convex structures \citep{Ester+1996,Von+2007}, they are not well suited to large-scale, high-dimensional settings. Both are sensitive to hyperparameters (e.g., $\epsilon$ and \textit{minPts} for DBSCAN \citep{Ester+1996}, and affinity-graph construction for Spectral Clustering \citep{Von+2007}) and are increasingly resource-intensive at scale. DBSCAN relies on costly neighborhood queries in high dimensions \citep{Gan+2015dbscan}, while Spectral Clustering requires an $N\times N$ affinity matrix \citep{Ng+2001spectral}, which is prohibitive for $N\sim10^{5}$. In our dataset of 99,806 samples with nearly $10^{3}$-dimensional features, both methods yielded trivial partitions dominated by a single large cluster, providing little morphological discrimination. Therefore, we exclude DBSCAN and Spectral Clustering from our framework. 
In the final hybrid clustering framework, we select K-means, Birch, and Agg as the core components. This choice is consistent with our previous series of experiments and analyses \citep{Zhou+2022, Song+2024, Yin+2025robust}. This combination achieves an optimal trade-off between classification accuracy, computational efficiency, and robustness. Our ensemble strategy proves particularly effective in handling the heterogeneity and scale of modern galaxy morphology datasets \citep{Zhou+2022,Yin+2025robust}. 

To further validate the effectiveness of the proposed hybrid clustering framework, we systematically compare independent clustering methods and the hybrid clustering method inspired by the Bagging technique. 
As shown in Table \ref{Related clustering results}, the clustering accuracy of the hybrid voting strategy is consistently higher than that achieved by individual algorithms. 
This result underscores the stability and consistency of the ensemble method in producing reliable morphological groupings. In our evaluation, the reference labels for computing the clustering accuracy are the cluster assignments produced by each clustering method, which are treated as pseudo-labels. These pseudo-labels are obtained by clustering AlexNet-extracted features that are reduced to 300 dimensions through UMAP. We then train a supervised classifier (GoogLeNet; \citealt{szegedy+2015}) on the preprocessed galaxy images using these pseudo-labels, and evaluate its prediction accuracy on a held-out test split (train/test = 9:1). This accuracy serves as a proxy metric to assess the effectiveness of the clustering results.

Although approximately 40\% of the samples are excluded during the consensus voting process due to label disagreement across base classifiers, the retained subset forms high-purity, high-confidence clusters with substantially lower ambiguity. These high-quality clusters serve as ready-to-use assets in downstream tasks. They are especially valuable for generating pre-labeled datasets with high annotation confidence, therefore, helping identify rare galaxy types in large-scale surveys and supporting training supervised models when annotated samples are scarce. Although the hybrid voting strategy reduces the effective training set size, the resulting improvement in label reliability better matches the requirements of high-precision classification. Importantly, objects excluded at this stage are subsequently reintroduced in the SML step of \texttt{USmorph} (see Fig.2 of \citealt{Yin+2025robust}), yielding morphology classifications for the full sample. This trade-off between coverage and purity demonstrates the effectiveness and utility of our hybrid clustering framework for galaxy morphology analysis.

\begin{table}[htbp]
\centering
\caption{Performance comparison of our method with different numbers of clusters.}
\small
\renewcommand{\arraystretch}{1}
\setlength{\tabcolsep}{10pt}
\begin{tabular}{|c|c|c|c|}
\hline
\textbf{Cluster} & \textbf{Method} & \textbf{Acc (\%)} & \textbf{Reject (\%)} \\
\hline
\multirow{4}{*}{8}
& K-means  & 86.1 & 0    \\ \hhline{|~|---|}
& Agg     & 88.4 & 0    \\ \hhline{|~|---|}
& Birch   & 84.5 & 0    \\ \hhline{|~|---|}
& Bagging & 95.4 & 42.7 \\ \hline
\multirow{4}{*}{12}
& K-means  & 91.3 & 0    \\ \hhline{|~|---|}
& Agg     & 88.6 & 0    \\ \hhline{|~|---|}
& Birch   & 80.1 & 0    \\ \hhline{|~|---|}
& Bagging & 96.2 & 40.2 \\ \hline
\multirow{4}{*}{16}
& K-means  & 92.1 & 0    \\ \hhline{|~|---|}
& Agg     & 90.1 & 0    \\ \hhline{|~|---|}
& Birch   & 90.7 & 0    \\ \hhline{|~|---|}
& Bagging & 97.5 & 40.3 \\ \hline
\multirow{4}{*}{20}
& K-means  & 91.0 & 0    \\ \hhline{|~|---|}
& Agg     & 90.0 & 0    \\ \hhline{|~|---|}
& Birch   & 88.0 & 0    \\ \hhline{|~|---|}
& Bagging & 94.6 & 41.6 \\ \hline
\end{tabular}
\par\vspace{2mm}
\noindent \footnotesize\justifying{\textit{Note:} Accuracy (Acc) marks the clustering classification accuracy, and Reject denotes the percentage of ambiguous or low-confidence samples excluded by the ensemble voting process. Bagging is employed as an ensemble voting framework that combines K-means, Agglomerative clustering, and Birch by aggregating their predictions through majority voting, with low-confidence samples rejected based on consensus.}
\label{Related clustering results}
\end{table}

\subsubsection{Cluster Number}

In unsupervised galaxy morphology classification, the choice of the number of clusters is critical for both model performance and the efficiency of subsequent human annotation. 
An appropriate cluster count shall balance between annotation workload and differentiation ability: excessive clusters increase the burden of manual inspection and labeling, whereas insufficient clusters may oversimplify the intrinsic morphological diversity, leading to inadequate separation between distinct galaxy types.

To investigate the influence of cluster number on classification performance, we conduct a series of experiments on a publicly available galaxy imaging dataset encompassing diverse morphological classes (e.g., ellipticals, spirals, irregulars, and mergers). As shown in Table \ref{Related clustering results}, clustering accuracy generally improves with an increasing number of clusters, as finer partitioning better captures subtle morphological variations. 
However, beyond an optimal point, the accuracy improves only marginally and may decline in some cases.
Two factors are proposed to be blamed for this decline: (1) an excessive number of clusters leads to over-segmentation in the feature space, producing small and unstable clusters that amplify noise instead of meaningful morphological differences; and (2) the increased number of clusters complicates manual inspection and labeling, introducing subjective bias and inconsistency that further degrade the overall clustering quality.
Our analysis suggests that setting the cluster number to 16 achieves an optimal balance between accuracy, stability, and interpretability. At this configuration, the clustering yields clear separation among major morphological populations, while the manual inspection of 16 representative cluster centroids can be efficiently completed by experts within minutes, without significant cognitive load.

It is noteworthy that the optimal number of clusters may depend on the scientific objectives and operational priorities. 
A moderate cluster count is preferable to ensure both accuracy and efficiency for the rapid processing of millions of galaxies in large-scale surveys conducted by next-generation telescopes. 
In contrast, a larger number of clusters is suggested even at the cost of higher complexity when attempting to uncover finer substructures, such as morphological transitions or rare intermediate forms. 
In this study, the unsupervised clustering step primarily serves to construct a high-confidence training dataset by grouping galaxies with consistent morphological characteristics. This refined dataset is then used to train a convolutional neural network (CNN), enabling accurate and efficient large-scale classification of galaxy morphologies across extensive imaging surveys. 
Therefore, the choice of cluster number should be driven by a synthesis of quantitative performance metrics, the desired level of morphological detail, and the downstream application requirements, balancing scientific interpretability with computational practicality.

\section{RESULTS AND DISCUSSION}\label{sec:4}

This section aims to validate the effectiveness of the proposed galaxy morphology classification framework. 
We first configure each component of the framework to its optimal settings.
We then present the classification results through t-SNE visualization and display both parameterized and non-parameterized measurement results. This comprehensive presentation demonstrates and substantiates the reliability of the classification framework.

\subsection{Optimal Configuration Selection}

Extensive experimental validation confirms that \texttt{USmorph} achieves optimal performance under specific configurations. The feature extraction module employs the ImageNet-pretrained AlexNet as the backbone encoder to fully capture hierarchical visual features in galaxy images. The dimensionality reduction stage utilizes the UMAP method to effectively preserve local and global structural information in the high-dimensional feature space. For the clustering analysis, three representative algorithms (K-means, Birch, and Agg) are used, with the number of clusters set to 16 to balance fine-grained classification and physical interpretability. 
Under this optimized configuration, unsupervised classification is performed on the selected galaxy dataset, yielding the clustering results shown in Table \ref{Related clustering results}. 
Galaxies are classified into five morphological types: spherical (SPH), early-type disk (ETD), late-type disk (LTD), irregular (IRR), and unclassified (UNC). SPHs typically exhibit a compact, smooth, and nearly symmetric light distribution. ETDs often show a prominent central bulge together with a disk component, but lack clear or coherent spiral arms and are generally less compact than SPHs. In contrast, LTDs are more extended and disk-dominated, frequently featuring well-defined spiral arms and a more diffuse light distribution. IRRs encompass a wide range of appearances, including asymmetric or disturbed structures, and may show features suggestive of past or ongoing interactions/mergers. UNCs usually correspond to low–signal-to-noise (S/N) cases and remain difficult to interpret morphologically: their faint surface-brightness profiles and limited structural detail hinder reliable placement within our classification scheme.

\begin{figure*}    
\includegraphics[width=1\textwidth]{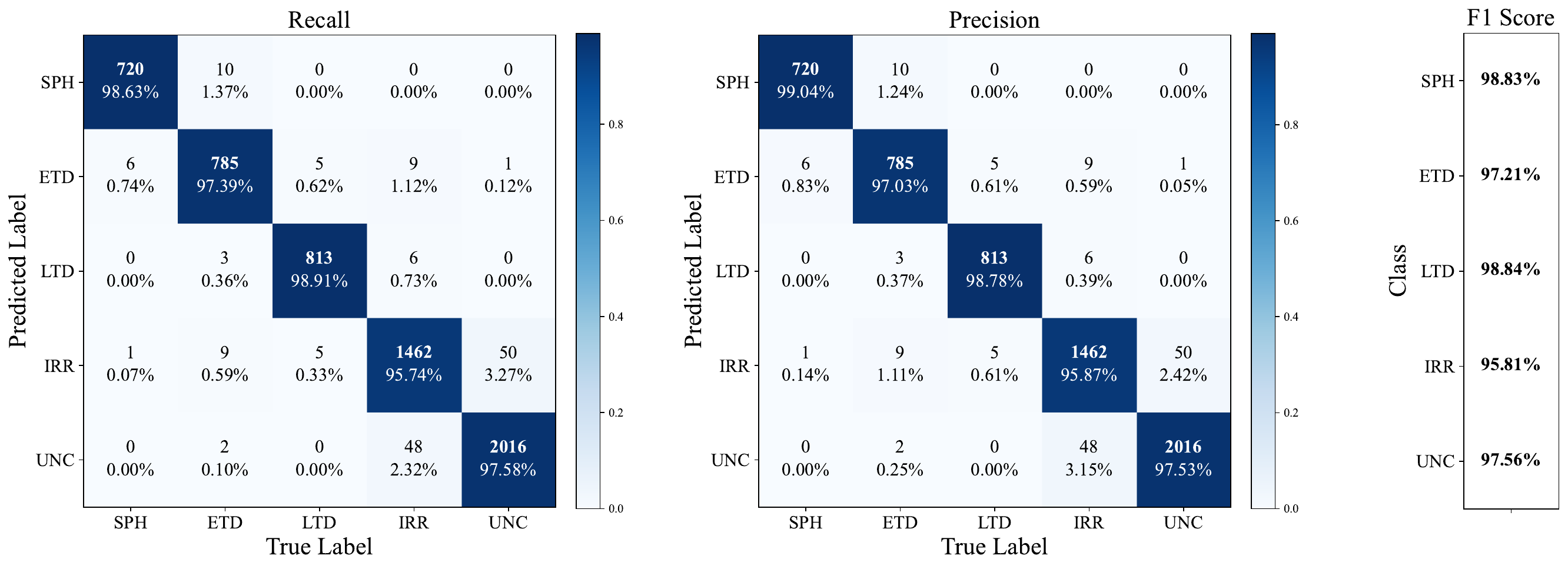}
\caption{The \textbf{left} and \textbf{middle} panels display the recall and precision rates for five galaxy classes, with overall recall and precision exceeding 97\%. The \textbf{right} panel presents the F1 scores for each galaxy class, demonstrating the UML framework's effectiveness in distinguishing between different galaxy types.}
\label{fig:7}
\end{figure*}

We then evaluate the precision, recall, and F1-score for each galaxy type. These metrics are defined as follows:
\begin{equation}
\mathrm{Precision}=\frac{TP}{TP+FP},
\end{equation}
\begin{equation}
\mathrm{Recall}=\frac{TP}{TP+FN},
\end{equation}
\begin{equation}
\mathrm{F1}=2\times\frac{\mathrm{Precision}\times\mathrm{Recall}}{\mathrm{Precision}+\mathrm{Recall}},
\end{equation}
where $TP$, $FP$, and $FN$ denote the numbers of true positives, false positives, and false negatives, respectively, computed in a one-vs-rest manner for each galaxy type. 
Here, the reference labels are the proxy morphological labels obtained by mapping the UML-generated clusters to morphological categories via manual visual inspection. We train GoogLeNet on the processed galaxy images using a 9:1 train/test split, and report precision, recall, and F1-score on the held-out test set (Figure~\ref{fig:7}). All three metrics exceed 95\% for every galaxy type, indicating that the UML-derived proxy labels are sufficiently consistent with the image morphology to enable large-scale supervised morphological classification.

\begin{figure*}    
\includegraphics[width=1\textwidth]{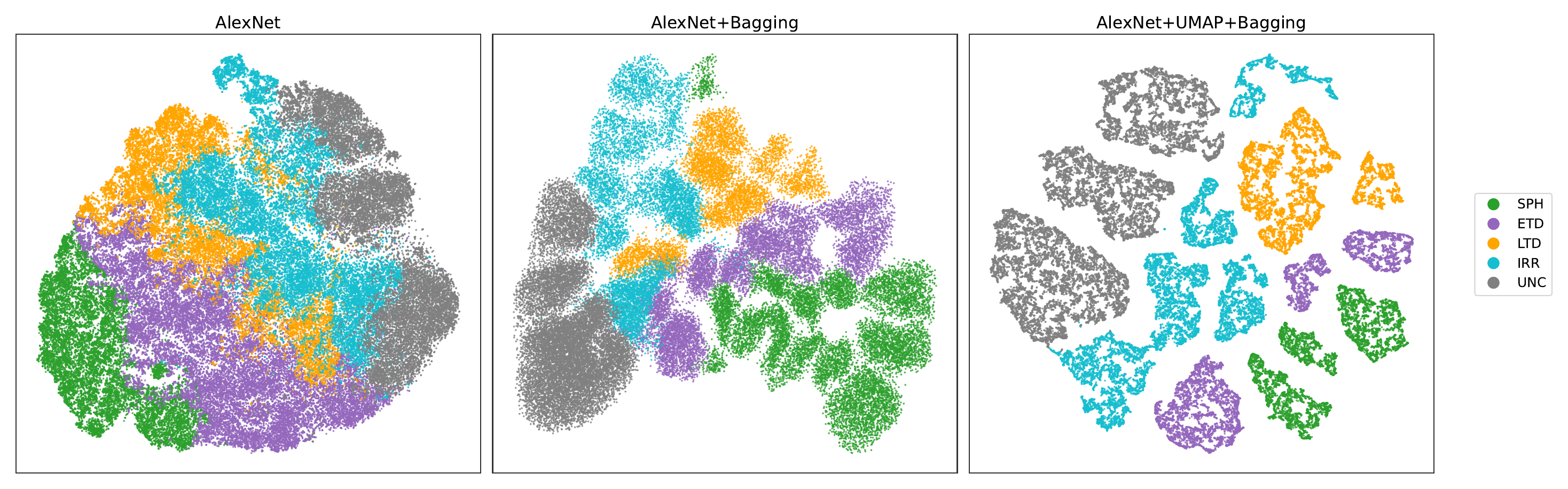}
\caption{Visualization of galaxy morphological classification results in the reduced feature space using different deep learning frameworks. From left to right, we show the results of: (1) AlexNet alone, (2) AlexNet combined with Bagging, and (3) AlexNet integrated with UMAP dimensionality reduction and Bagging. Each color represents a distinct morphological class: green for SPH, purple for ETD, orange for LTD, blue for IRR, and gray for UNC. The better separation between classes in the third panel indicates enhanced clustering performance through dimensionality reduction and ensemble methods.}
\label{fig:8}
\end{figure*}
\begin{figure*}    
\includegraphics[width=1\textwidth]{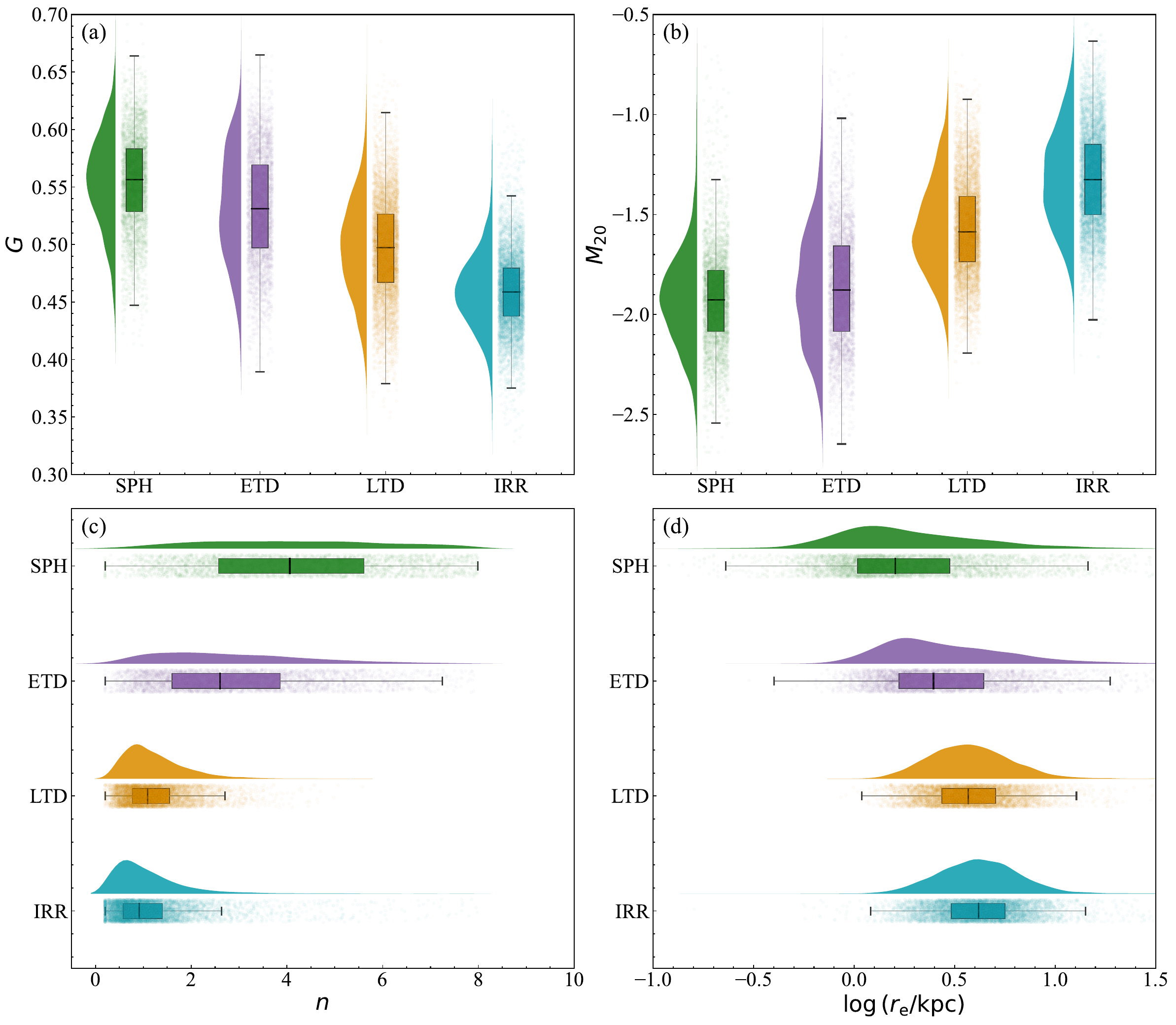}
\caption{Distribution of key morphological parameters for different galaxy classes identified by the unsupervised classification pipeline. Panels show the results of: 
    \textbf{(a)} Gini coefficient $G$, 
    \textbf{(b) }concentration index $M_{20}$, 
    \textbf{(c)} Sérsic index $n$, and 
    \textbf{(d)} effective radius $r_e$.
    Each panel includes violin plots with overlaid boxplots, where the black solid line represents the median, the thick black bar indicates the interquartile range (IQR), and the thin whiskers extend to 1.5$\times$IQR. 
    Galaxy types are color-coded as follows: SPH in green, ETD in purple, LTD in orange, and IRR in blue. 
    The distributions reveal distinct trends among galaxy types, supporting the physical coherence of the classification.
}
\label{fig:9}
\end{figure*}

\subsection{ Overall Morphological Classification Results and t-SNE }
 
We employ t-distributed stochastic neighbor embedding (t-SNE; \citealt{van+2008}) to visualize the clustering results in a two-dimensional space derived from the high-dimensional feature space. t-SNE effectively preserves local data structures in high dimensions, enabling clear visualization of distribution relationships between categories in low dimensions. Additionally, this paper compares t-SNE visualizations obtained using a single clustering algorithm, a single clustering algorithm combined with UMAP dimensionality reduction, and the complete framework (incorporating UMAP dimensionality reduction and the Bagging multi-voting strategy). 
The results are shown in Figure \ref{fig:8}.
Visualizations under different configurations demonstrate that incorporating UMAP dimensionality reduction and ensemble clustering strategies significantly enhances category separability and clustering compactness, validating the proposed method's effectiveness in feature representation and classification stability.

\subsection{Comparisons with Galaxy Properties}

To assess the consistency between the classification results and the known structural properties of galaxies, we examine the morphological classifications in relation to key physical parameters, as shown in Figure~\ref{fig:9}. 
The analysis is restricted to massive galaxies with stellar masses $M_{*} > 10^{9}~M_{\sun}$ to ensure robust and reliable parameter measurements.
UNC galaxies are excluded from this analysis since their low S/Ns hinder accurate determination of morphological parameters. 
In Figure~\ref{fig:9}, galaxies are color-coded by class: SPH in green, ETD in purple, LTD in orange, and IRR in blue. 

Panel \textbf{(a)} shows the Gini coefficient ($G$) distribution of four morphological classes: SPH, ETD, LTD, and IRR. The Gini coefficient quantifies the non-uniformity of the surface brightness distribution, with higher values indicating stronger concentration or clumpiness of light \citep{Lotz+2004,Lotz+2008}. SPH galaxies exhibit the highest $G$ values, consistent with their compact bulge-dominated structures, while IRR galaxies show lower $G$ values due to their irregular, diffuse morphologies.

Panel \textbf{(b)} presents the distribution of the $M_{20}$ parameter, which measures the normalized second-order moment of the brightest 20\% of a galaxy’s flux (\citealt{Lotz+2004, Lotz+2008}). More negative $M_{20}$ values correspond to a higher central light concentration. The distribution shows that SPH galaxies have the most negative $M_{20}$ values, indicating higher concentration.  In contrast, LTD and IRR galaxies tend to have higher $M_{20}$, reflecting less concentrated light distributions.

Panel \textbf{(c)} displays raincloud plots of the Sérsic index ($n$), which characterizes the steepness of the galaxy’s radial surface brightness profile (\citealt{Zhou+2022, Song+2024}). Higher $n$ values signify a more prominent bulge component, whereas lower $n$ values correspond to disk-dominated or irregular systems. The median $n$ increases systematically from IRR to SPH galaxies, revealing a clear morphological sequence.

Panel \textbf{(d)} shows the distribution of the effective radius ($r_e$), defined as the physical radius enclosing half of the galaxy’s total light. SPH galaxies generally possess smaller $r_e$, corresponding to compact, centrally concentrated structures, while LTD and IRR galaxies display larger $r_e$ values, which are consistent with their more extended morphologies.

All parameters in Figure \ref{fig:9} demonstrate that our morphological classification aligns with the expected morphological sequence: from IRR to SPH galaxies, (a) the light distribution becomes increasingly concentrated, (b) $M_{20}$ decreases, (c) Sérsic $n$ rises, and (d) $r_e$ shrinks, indicating a transition from extended star-forming systems to compact, quiescent spheroids.

\section{Summary and outlooks} \label{sec:6}
This paper presents a systematic analysis of the robustness of the unsupervised machine learning (UML) component within the hybrid framework \texttt{USmorph}, which is designed for the automated morphology classification of galaxies in large-scale surveys. The framework has been validated on a sample of nearly 100,000 I-band galaxies ($0.2 < z < 1.2$, $I_{\mathrm{mag}} < 25$) from the COSMOS field, demonstrating its effectiveness and reliability in morphological characterization.

We test different configurations of key components within the \texttt{USmorph} framework to reach an optimal performance:
\begin{enumerate}[label=(\arabic*)]

\item In terms of feature extraction, we compare the five mainstream encoders and find that the AlexNet architecture is best at capturing hierarchical morphological features from galaxy images.

\item For dimensionality reduction, UMAP is identified as the optimal method, effectively preserving both local and global structures of the high-dimensional feature space while balancing computational efficiency and clustering accuracy. 

\item Regarding clustering algorithms, K-means, Birch, and Agg exhibit consistent stability and strong performance across multiple experiments, making them the preferred choices within our framework. A Bagging-based multi-clustering ensemble strategy is further employed, which significantly enhances label purity and the robustness of consensus clustering compared to single clustering approaches.

\item The number of clusters in the Bagging-based multi-clustering step is set to $K=16$, achieving a practical balance between classification granularity and annotation efficiency. 
\end{enumerate}
This configuration enables clear separation of major galaxy morphologies. We further confirm that the morphology classification results align with galaxy evolution theory, showing physically plausible distributions of different types in the parameter space. 
Therefore, \texttt{USmorph} provides a scalable foundation for fine-grained identification and further investigation of rare or peculiar objects, such as tidal tails, gravitational lenses, and little red dots.

In future work, we aim to enhance the scalability and generalization of the automated galaxy morphology classification framework by integrating advanced deep representation learning with domain adaptation techniques. We will explore self-supervised and contrastive learning paradigms to learn robust morphological representations from large-scale unlabeled imaging data. Furthermore, we will investigate unsupervised domain adaptation (UDA) and domain-invariant feature learning to mitigate distribution shifts across different surveys, such as those arising from variations in resolution, depth, and point-spread function. 
This is crucial for ensuring that models trained on data from one telescope or bandpass can be reliably transferred to another. 
Our ultimate goal is to develop morphology-aware models that generalize across observational depth, spatial resolution, and wavelength. The enhanced framework will be applied to upcoming multi-band deep-field surveys, particularly the imaging data from the Chinese Space Station Telescope (CSST). This application is anticipated to enable large-sample morphological analysis with high statistical precision and facilitate the identification and detailed characterization of rare or peculiar structural features in galaxy evolution, such as tidal tails, stellar shells, and merger remnants.

\section*{Data Availability}
\textbf{Our code, dataset, and trained model weights can be found at \url{https://github.com/IAAA-246011/USmorph_2.0}.}

\begin{acknowledgements}
This work is supported by the National Natural Science Foundation of China (NSFC) (grant Nos. 12233008 and 12573012), the National Key R\&D Program of China (grant No. 2023YFA1608100), the Strategic Priority Research Program of the Chinese Academy of Sciences (grant No. XDB0550200), the Cyrus Chun Ying Tang Foundations, and the 111 Project for “Observational and Theoretical Research on Dark Matter and Dark Energy” (B23042), and the China Manned Space Program with grant No. CMS-CSST-2025-A04. The numerical calculations in this paper have been done on the platform of the High Performance Computing of Anqing Normal University.

\end{acknowledgements}

\bibliography{ref}

\end{document}